# Acoustic spin-1 Weyl semimetal

Weiyin Deng[1*], Xueqin Huang[1*], Jiuyang Lu[1], Feng Li[1†], Jiahong Ma[1], Shuqi Chen[3†], Zhengyou Liu[2,4†]

[1]School of Physics and Optoelectronics, South China University of Technology, Guangzhou, Guangdong 510640, China

[2]Key Laboratory of Artificial Micro- and Nanostructures of Ministry of Education and School of Physics and Technology, Wuhan University, Wuhan 430072, China

[3]Key Laboratory of Weak Light Nonlinear Photonics of Ministry of Education, School of Physics, Nankai University, Tianjin 300071, China

[4]Institute for Advanced Studies, Wuhan University, Wuhan 430072, China

*W.D. and X.H contributed equally to this work

†Corresponding author. Email: phlifeng@scut.edu.cn; schen@nankai.edu.cn; zyliu@whu.edu.cn

**Abstract:** Topological semimetal, hosting spin-1 Weyl point beyond Dirac and Weyl points, has attracted a great deal of attention. However, the spin-1 Weyl semimetal, which possesses exclusively the spin-1 Weyl points in a clean frequency window, without shadowed by any other nodal points, is yet to be discovered. Here, we report for the first time a spin-1 Weyl semimetal in a phononic crystal. Its spin-1 Weyl points, touched by two linear dispersions and an additional flat band, carry monopole charges $(-2, 0, 2)$ or $(2, 0, -2)$ for the three bands from bottom to top, and result in double Fermi arcs existing both between the 1st and 2nd bands, as well as between the 2nd and 3rd bands. We further observe robust propagation against the multiple joints and topological negative refraction of acoustic surface arc wave. Our results pave the way to explore on the macroscopic scale the exotic properties of the spin-1 Weyl physics.

## Introduction

The discovery of Dirac and Weyl semimetals [1-8], featured with fourfold and twofold linear crossing points in the band structures, has opened up a new research field known as topological semimetals [9, 10]. The low-energy excitations near the Dirac point and Weyl point (WP) are described by the Dirac and Weyl equations, and thus, behave like Dirac and Weyl fermions, i.e., the relativistic spin-1/2 fermions in quantum field theory. Unlike the fermions in high-energy theories, the excitations in crystals are protected by the rich symmetries of the space group, rather than the Poincaré symmetry, and give rise to new physical phenomena, such as type-II Weyl semimetals [11], nodal line semimetals [12] and spin-1 Weyl semimetals [13]. The spin-1 Weyl semimetals, which host exclusively the spin-1 WPs in a frequency window without shadowed by any other nodal points, have been proposed in electronic and cold atom materials recently [13-15]. However, the spin-1 Weyl semimetal is yet to be realized.

The spin-1 WPs are threefold degenerate points touched by two linear dispersions and an additional flat band, and carry topological charges $(-2, 0, 2)$ or $(2, 0, -2)$ [13]. The spin-1 WPs with nontrivial topology or chirality result in numerous intriguing properties, such as the double Fermi arcs [13, 14], chiral anomaly [13], the quantized circular photogalvanic effect [16-18], chiral optical response [18, 19], and novel Imbert-Fedorov shift [20] and so on. The spin-1 WP coexisting with other chiral multifold points, have been attracting significant attentions in both theoretical predictions [16, 21-24] and experimental observations [25-31]. Although the spin-1 WPs have been reported in Refs. [25-31], the topological charges of spin-1 WPs were not completely verified. Only the topological charge $-2$ of the 1$^{st}$ band of the spin-1 WP can be determined by the double Fermi arcs. It is necessary to realize a real spin-1 Weyl semimetal with pure spin-1 WPs, to explore the topological physics of spin-1 WPs.

In this work, we report the realization of spin-1 Weyl semimetal in a three-dimensional (3D) phononic crystal (PC). Compared with electronic materials, the unique advantage of PCs lies in their macroscopic scale and flexibility of fabrication.

In particular, as there are only two spin-1 WPs near a frequency surface in our design, the separation between the WPs can be sufficiently wide, so that the WPs and the associated surface arcs, which are similar to the Fermi arcs in electronic systems, can be easily accessed in experiments. To be specific, we first search the spin-1 WPs in a layer-stacking Lieb lattice, and show the phase diagram and topologically protected surface states. Then, we design a 3D PC corresponding to this lattice model and present the experimental observation of spin-1 WPs and surface arcs. The theoretical and experimental results are in good agreement.

## Results

### Tight-binding model

To illustrate how to derive spin-1 WPs in this work, we construct a tight-binding model based on a layer-stacking Lieb lattice with a 3-site unit cell, denoted by A (red sphere), B (yellow sphere), and C (blue sphere) in the upper panel of Fig. 1a. The nearest-neighbor hoppings of the intralayer in the XY plane are $t_0$, and the interlayer couplings (dashed lines) are chiral with strength $t_z$. On the bases of the sublattices A-C, the Bloch Hamiltonian of this model can be described by

$$H = \begin{pmatrix} 0 & 2t_0 \cos(k_x a/2) & 2t_0 \cos(k_y a/2) \\ 2t_0 \cos(k_x a/2) & 0 & t_{z1} \cos(k_z h) - it_{z2} \sin(k_z h) \\ 2t_0 \cos(k_y a/2) & t_{z1} \cos(k_z h) + it_{z2} \sin(k_z h) & 0 \end{pmatrix}, (1)$$

where $t_{z1} = 4t_z \cos(k_x a/2) \cos(k_y a/2)$, $t_{z2} = 4t_z \sin(k_x a/2) \sin(k_y a/2)$, and $t_0, t_z < 0$. $a$ and $h$ are the lattice constants in the $x$ (or $y$) and $z$ directions, respectively. Diagonalizing the Hamiltonian (1) yields a pair of spin-1 WPs with zero energy located at points $M = (\pi/a, \pi/a, 0)$ and $R = (\pi/a, \pi/a, \pi/h)$ in the first Brillouin zone (BZ), as shown by the green and purple spheres in the left panel of Fig. 1b. These two spin-1 WPs at the M and R points are determined by three coexisting chiral symmetries along three mutually orthogonal lines (Supplementary Material I). In all momentum directions, the threefold degenerate point consists of two linear dispersions and an additional flat band, as shown in Figs. 1c and 1d. To explore the underlying physics of the threefold degenerate points, we derive their low-energy effective excitations, and find that they have the form $\boldsymbol{k}\cdot\boldsymbol{S}$, where $\boldsymbol{S}$ is the vector of the spin-1 matrix representing the sub-lattice pseudospin, as described in Supplementary Material II. These threefold degenerate points are spin-1 WPs, since they are spin-1 generalizations of WPs. The emergence of spin-1 WPs can be interpreted as follows: The single layer, as a Lieb lattice, can support a two-dimensional (2D) threefold degenerate point at $(k_x, k_y) = (\pi/a, \pi/a)$ with a Hamiltonian of the form $k_x S_x + k_y S_y$; meanwhile, the chiral interlayer coupling contributes to the third term of the Hamiltonian $k_z S_z$ near $k_z = 0$ or $\pi/h$ to generate the spin-1 WPs.

We now study the topological properties of the system. We first calculate the monopole charges of the spin-1 WPs, and obtain $C_\text{M} = (2, 0, -2)$ and $C_\text{R} = (-2, 0, 2)$ (Supplementary Material III). The spin-1 WPs behave as monopoles of Berry flux in momentum space, which are twice those of the WPs with spin-1/2. By considering $k_z$ as a parameter, the Chern numbers $C(k_z)$ of the three bands are plotted in the right panel of Fig. 1b. A more complete phase diagram is shown in Supplementary Material IV. The monopole charges of the spin-1 WPs are consistent with the $k_z$-dependent Chern numbers $C(k_z)$, for instance, the charge for the first band at M is $C(k_z = 0^+) - C(k_z = 0^-) = 2$. Choosing the open boundaries in the *x* direction in the bottom panel of Fig. 1a, we plot the surface arcs between the 1[st] and 2[nd] bands on the normal (N) and hollow (H) surfaces shown in Figs. 1e and 1f, respectively. Two surface arcs connect the projections of two oppositely charged spin-1 WPs, coexisting with bulk states of the second band near the spin-1 WPs. The surface arcs between the 2[nd] and 3[rd] bands and the surface state dispersions are displayed in Supplementary Material V. The double surface arcs, both existing between the 1[st] and 2[nd] bands, and between the 2[nd] and 3[rd] bands, provide solid evidences for the topological charges of the spin-1 WPs. The illustration of the spin-1 WPs associated surface arcs is shown in Supplementary Material VI.

## Acoustic spin-1 Weyl points

Let us consider the real PC for realizing spin-1 Weyl semimetal for acoustic waves. As shown in Fig. 2a, the PC sample is a layer-stacking structure fabricated by 3D printing. The unit cell, shown in Fig. 2b, contains three non-equivalent cavities linked up with tubes, where the grey areas represent hard boundaries, and the green areas denote periodic boundaries. The cavities can be viewed roughly as the lattice sites, while the tubes provide the hopping parameters. The intralayer tubes are set to be the same size to give rise to equal couplings in the XY plane, corresponding to the Lieb lattice. The interlayer tubes are chiral to connect the different cavities, which induce an effective gauge flux [7].

The structure possesses the P4 (No. 75) and time-reversal symmetries, which do not have 3D irreducible representations at the high symmetry points. So the spin-1 WPs are only able to be induced by the accidental degeneracy of a one-dimensional and a 2D irreducible representations at M and R. The size $l_2$ of one cavity is adjusted and different to size $l_1$ of the other cavities (Supplementary Material VII). At M, if $l_2$ is smaller than 3mm, the frequency of the double degenerate state is higher than that of the single state; while $l_2$ is larger than 3mm, the frequency of the double degenerate state is lower than that of the single state. The band inversion of the single state and the double degenerate state with different $l_2$ guarantees that there must exist a threefold degenerate state for $l_2 = 3\text{mm}$. Actually, the spin-1 WPs are easily generated by accidental degeneracy in real phononic crystals, since the acoustic sample is closely a one-to-one mapping of the tight-binding model with cavities mapping to atoms and tubes to bonds. The existence of the spin-1 WPs guarantees that the Hamiltonian have uniform on-site energies. Therefore, the real PC could be well modelled by the tight-binding Hamiltonian (Supplementary Material VIII).

To confirm the existence of the pair of acoustic spin-1 WPs shown in Fig. 1b, we calculate and measure the 2D band structures for fixed $k_z = 0$ and $k_z = \pi/h$ in Figs. 2c and 2e, respectively. The color maps represent the experimental dispersions expressed in terms of the average intensity of the Bloch states, while the white circles represent the simulated values obtained from full-wave simulations (see Materials and Methods). Threefold degenerate points at the M and R points are clearly observed at approximately the same frequency 7.5kHz. The finer structure near the Γ point below the second band in Fig. 2c may result from the multiple reflection by the surfaces of the finite sample. In Fig. 2d, the dispersion on the plane $k_z = 0.5\pi/h$ opens up two gaps between the three bands at the $\bar{M}$ point, due to effective time-reversal symmetry breaking by the synthetic gauge flux. The simulated and experimental results for the band dispersions along the $k_z$ direction are shown in Fig. 2f, corresponding to the band structure in Fig. 1d. One can find that the dispersions near the threefold degenerate points are crossings of two linear bands and an additional flat band, clearly indicating that the two threefold degenerate points are spin-1 WPs in the 3D band structure.

**Acoustic surface arc waves**

Theoretical studies predict that the non-zero Chern numbers $C(k_z)$ can lead to a pair of gapless surface states for a ribbon structure. Experimentally, the surface state dispersions can be obtained by measuring and Fourier transforming the surface acoustic fields, which are shown in Supplementary Material IX. Given $k_z = 0.5\pi/h$, the surface state dispersions along the $k_y$ direction are plotted in Figs. 3a and 3b, for the N and H surfaces, respectively. The surface state dispersions between the 1st and 2nd bands, as well as between the 2nd and 3rd bands are observed, resulting from the same topology for $C(k_z) = 0$ of the second band. Compared with the dispersions of surface state in the tight-binding model, one can see that the results for the PC are consistent with the model.

For a given excitation frequency, the surface states in the PC will trace out trajectories to connect the two oppositely charged spin-1 WPs and form surface arcs. At $f = 7$kHz, the surface arcs between the 1st and 2nd bands on the N and H surfaces are shown in Figs. 3c and 3d, respectively. The surface arcs connecting the pair of spin-1 WPs exhibit two branches, and exist in $k_z \in (0, \pi/h)$ and $k_z \in (-\pi/h, 0)$. In addition, the surface arcs near the spin-1 WPs merge with the projections of the bulk states. These features are well consistent with the theoretical results in Figs. 1e and 1f. The similar features of surface arcs at $f = 9\text{kHz}$ are also observed between the 2$^{\text{nd}}$ and 3$^{\text{rd}}$ bands, as shown in Figs. 3e and 3f. The double surface arcs, observed between the 1$^{\text{st}}$ and 2$^{\text{nd}}$ bands, as well as between the 2$^{\text{nd}}$ and 3$^{\text{rd}}$ bands, verified, directly and completely, the topological charges of the spin-1 WPs.

**Topological reflection-free propagations**

The surface arcs with nonclosed nature can give rise to intriguing topological phenomena, including the anomalous quantum oscillations [32, 33], 3D quantum Hall effect [34, 35], and topological negative refraction [36]. In Fig. 4, we provide two experiments visualizing unusual topological reflection-free propagations of surface arc wave. The first experiment demonstrates the robust propagation of the surface arc wave against the multiple joints of adjacent facets, as shown in Figs. 4a and 4b. The second one displays the topological negative refraction, as shown in Figs. 4c-4e. One can see that the surface arc waves can propagate only in an anticlockwise manner along the boundary, and do not reflect or scatter, because of the non-closed surface arc configuration (Fig. 4d). It is worth pointing out that these topological propagations can occur at the frequencies not only between the first and second bands, in the meantime also between the second and third bands, as shown in Supplementary Material X. The attenuations of the surface arc waves during propagation due to the absorption of air and the decaying behaviours of the surface arc waves along the perpendicular direction are shown in Supplementary Material XI.

**Conclusions**

We have realized an acoustic spin-1 Weyl semimetal, which opens up an avenue to explore new topological physics in 3D acoustics, besides spin-1/2 Weyl semimetals [7, 8, 36-38]. Our system, as a real spin-1 Weyl semimetal, is essential for exploring the topological properties of spin-1 WPs, including the surface arc states and chiral anomaly [13]. The robust propagation against the multiple joints of adjacent facets and topological negative refraction of surface arc wave may serve as a basis for designing innovative acoustic devices. As possessing slow group velocity and high density of state, the flat bands of spin-1 WPs may be used to localize or freeze acoustic waves [39-41], and enhance local fields [42-44]. Moreover, the layer-stacking method in our work can extend to other periodic structures, including electronic, photonic and cold atom systems. Finally, the chiral Landau levels induced by a pseudo-magnetic field in our spin-1 Weyl semimetal is of great interest, which can be implemented through the inhomogeneous potentials [38].

## Supplementary Data

Supplementary data are available online.

## Funding

This work is supported by the National Natural Science Foundation of China (Grants Nos. 11890701, 11604102, 11704128, 11774275, 11804101, 11974005, and 11974120); the National Key R&D Program of China (Grant No. 2018FYA0305800); Guangdong Innovative and Entrepreneurial Research Team Program (Grant No. 2016ZT06C594), and the Fundamental Research Funds for the Central Universities (Grants No. 2018MS93, No. 2019JQ07, and No. 2019ZD49).

# References

1. Wan X, Turner AM, Vishwanath A, Savrasov SY. Topological semimetal and Fermi-arc surface states in the electronic structure of pyrochlore iridates. *Phys Rev B* 2011; **83**: 205101.

2. Wang Z et al. Dirac semimetal and topological phase transitions in $A_3Bi$ (A = Na, K, Rb). *Phys Rev B* 2012; **85**: 195320.

3. Liu ZK et al. Discovery of a three-dimensional topological Dirac semimetal, $Na_3Bi$. *Science* 2014; **343**: 864-7.

4. Lv BQ et al. Experimental discovery of Weyl semimetal TaAs. *Phys Rev X* 2015; **5**: 031013.

5. Xu SY et al. Discovery of a Weyl fermion semimetal and topological Fermi arcs. *Science* 2015; **349**: 613-7.

6. Lu L et al. Experimental observation of Weyl points. *Science* 2015; **349**: 622-4.

7. Xiao M et al. Synthetic gauge flux and Weyl points in acoustic systems. *Nat Phys* 2015; **11**: 920-4.

8. Li F et al. Weyl points and Fermi arcs in a chiral phononic crystal. *Nat Phys* 2018; **14**: 30-4.

9. Burkov AA. Topological semimetals. *Nat Mater* 2016; **15**: 1145-8.

10. Armitage NP, Mele EJ, Vishwanath A. Weyl and Dirac semimetals in three-dimensional solids. *Rev Mod Phys* 2018; **90**: 015001.

11. Soluyanov AA et al. Type-II Weyl semimetals. *Nature* 2015; **527**: 495-8.

12. Burkov AA, Hook MD, Balents L. Topological nodal semimetals. *Phys Rev B* 2011; **84**: 235126.

13. Bradlyn B et al. Beyond Dirac and Weyl fermions: Unconventional quasiparticles in conventional crystals. *Science* 2016; **353**: aaf5037.

14. Zhu YQ et al. Emergent spin-1 Maxwell fermions with a threefold degeneracy in optical lattices. *Phys Rev A* 2017; **96**: 033634.

15. Fulga IC, Fallani L, Burrello M. Geometrically protected triple-point crossings in an optical lattice. *Phys Rev B* 2018; **97**: 121402(R).

16. Chang G et al. Unconventional chiral fermions and large topological Fermi arcs in RhSi. *Phys Rev Lett* 2017; **119**: 206401.

17. Rees D et al. Quantized photocurrents in the chiral multifold Fermion system RhSi. arXiv:1902.03230 (2019).

18. Flicker F et al. Chiral optical response of multifold fermions. *Phys Rev B* 2018; **98**: 155145.

19. Sanchez-Martinez MA, Juan F, Grushin AG. Linear optical conductivity of chiral multifold fermions. *Phys Rev B* 2019; **99**: 155145.

20. Hao YR, Wang L, Yao DX. Imbert-Fedorov shift in pseudospin-N/2 semimetals and nodal-line semimetals. *Phys Rev B* 2019; **99**: 165406.

21. Manes JL. Existence of bulk chiral fermions and crystal symmetry. *Phys Rev B* 2012; **85**: 155118.

22. Tang P et al. Multiple types of topological fermions in transition metal silicides. *Phys Rev Lett* 2017; **119**: 206402.

23. Saba M et al. Group theoretical route to deterministic Weyl points in chiral photonic lattices. *Phys Rev Lett* 2017; **119**: 227401.

24. Zhang T et al. Double-Weyl phonons in transition-metal monosilicides. *Phys Rev Lett* 2018; **120**: 016401.

25. Bao S et al. Discovery of coexisting Dirac and triply degenerate magnons in a three-dimensional antiferromagnet. *Nat Commun* 2018; **9**: 2591.

26. Miao H et al. Observation of double Weyl phonons in parity-breaking FeSi, *Phys Rev Lett* 2018; **121**: 035302.

27. Takane D et al. Observation of chiral Fermions with a large topological charge and associated Fermi-arc surface states in CoSi. *Phys Rev Lett* 2019; **122**: 076402.

28. Rao Z et al. Observation of unconventional chiral fermions with long Fermi arcs in CoSi. *Nature* 2019; **567**: 496-9.

29. Sanchez DS et al. Topological chiral crystals with helicoid-arc quantum states. *Nature* 2019; **567**: 500-5.

30. Yang Y et al. Topological triply-degenerate point with double Fermi arcs. *Nat Phys* 2019; **15**: 645-9.

31. Schroter NBM et al. Chiral topological semimetal with multifold band crossings and long Fermi arcs. *Nat Phys* 2019; **15**: 759-65.

32. Potter AC, Kimchi I, Vishwanath A. Quantum oscillations from surface Fermi arcs in Weyl and Dirac semimetals. *Nat Common* 2014; **5**: 5161.

33. Moll PJW et al. Transport evidence for Fermi-arc-mediated chirality transfer in the Dirac semimetal $Cd_3As_2$. *Nature* 2016; **535**: 266-70.

34. Wang CM, Sun HP, Lu HZ, Xie XC. 3D Quantum Hall Effect of Fermi Arcs in Topological Semimetals. *Phys Rev Lett* 2017; **119**: 136806.

35. Zhang C et al. Quantum Hall effect based on Weyl orbits in $Cd_3As_2$. *Nature* 2018; **565**: 331-6.

36. He H et al. Topological negative refraction of surface acoustic waves in a Weyl phononic crystal. *Nature* 2018; **560**: 61-4.

37. Ge H et al. Experimental observation of acoustic Weyl points and topological surface states. *Phys Rev Appl* 2018; **10**: 014017.

38. Peri V et al. Axial-field-induced chiral channels in an acoustic Weyl system. *Nat Phys* 2019; **15**: 357-61.

39. Vicencio RA et al. Observation of localized states in Lieb photonic lattices. *Phys Rev Lett* 2015; **114**: 245503.

40. Mukherjee S et al. Observation of a localized flat-band state in a photonic Lieb lattice. *Phys Rev Lett* 2015; **114**: 245504.

41. Taie S et al. Coherent driving and freezing of bosonic matter wave in an optical Lieb lattice. *Sci Adv* 2015; **1**: e1500854.

42. Li Y et al. Enlargement of locally resonant sonic band gap by using composite plate-type acoustic metamaterial. *Phys Lett A* 2015; **379**: 412-6.

43. Leykam D, Flach S. Perspective: Photonic flatbands. *APL Photonics* 2018; **3**: 070901.

44. Longhi S. Photonic flat-band laser. *Opt Lett* 2019; **44**: 287-90.

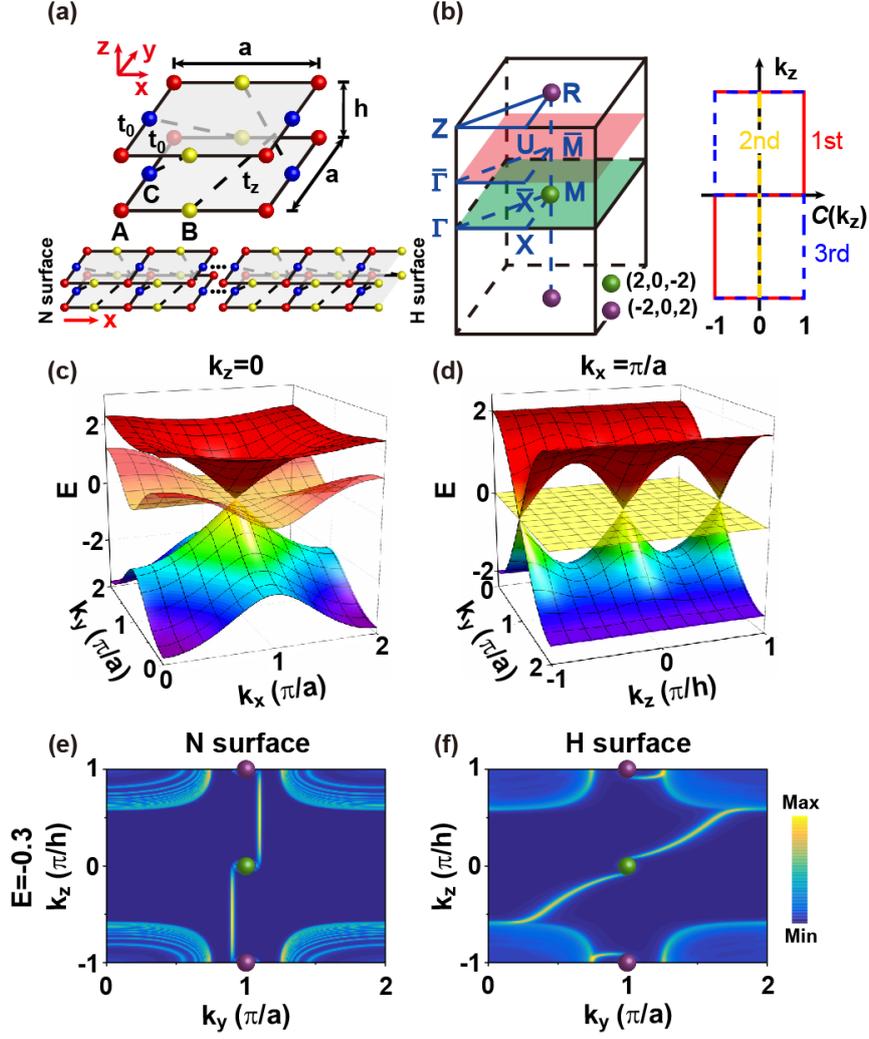

**Figure 1.** Bulk band structures and surface arcs of the spin-1 Weyl points for a lattice model. (a) Schematic of the unit cell and ribbon. Upper panel: Red, yellow, and blue spheres denote A, B, and C lattices in a unit cell. Lower panel: Ribbon for calculation of the surface state dispersions. Two different opposite surfaces are marked as N and H surfaces, respectively. (b) Left panel: The first BZ of the system. The green and purple spheres represent spin-1 WPs with topological charges $(2, 0, -2)$ and $(-2, 0, 2)$, respectively. Right panel: The Chern numbers of the three bands as a function of $k_z$. (c)-(d) The 3D bulk band structures with $k_z = 0$ and $k_x = \pi/a$, respectively. (e)-(f) The surface arcs between the 1$^{st}$ and 2$^{nd}$ bands for $E = -0.3$ on the N and H surfaces, respectively. The coupling parameters are taken to $t_0 = -1$ and $t_z = -0.3$.

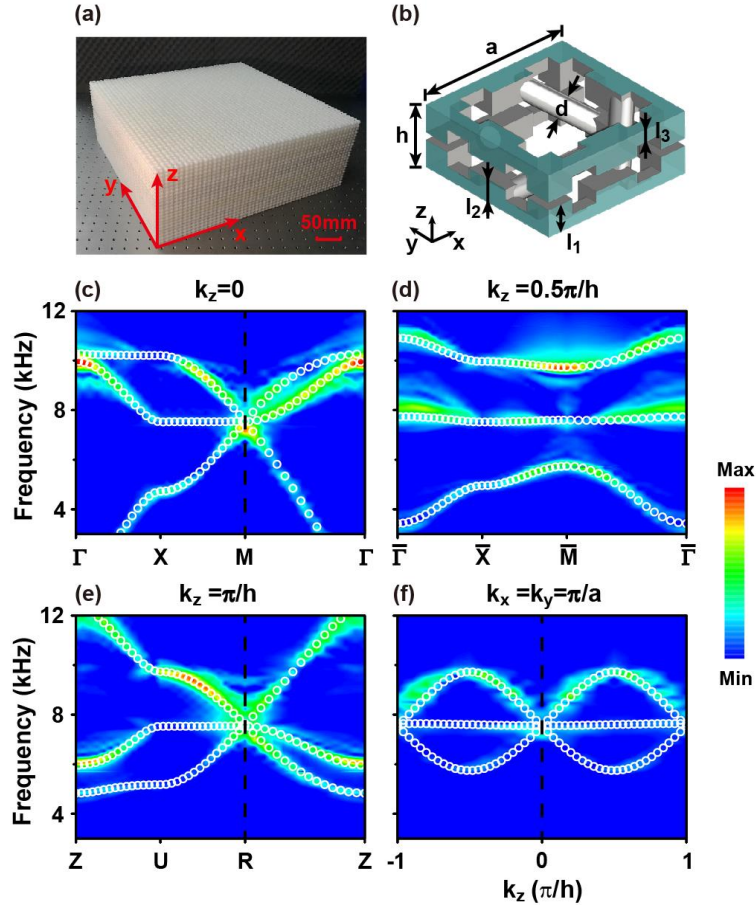

**Figure 2.** Experimental demonstration of the bulk band structures of the spin-1 Weyl points in a phononic crystal. (a) A photo of the 3D sample. (b) The unit cell of the phononic crystal. Here, $a = 20$mm, $h = 8$mm, $l_1 = 3.5$mm, $l_2 = 3$mm, $l_3 = 1.6$mm and $d = 2.72$mm. The periodic boundaries are applied in the *x*, *y* and *z* directions. The grey areas represent hard boundaries. (c)-(f) The bulk band structures of the three lowest modes in $k_z = 0$, $k_z = 0.5\,\pi/h$, $k_z = \pi/h$ and along the $k_z$ direction. The color maps represent the experimental data, while the white circles represent the full-wave simulation results.

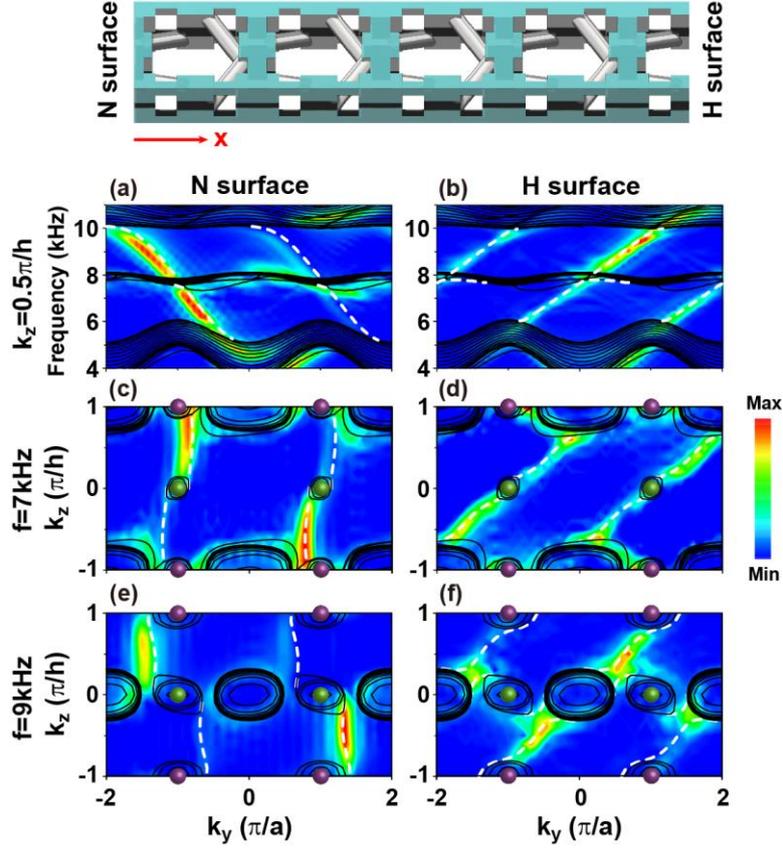

**Figure 3.** Dispersions of surface state and acoustic surface arcs. The top picture shows a schematic of the ribbon for full-wave simulations of surface state dispersions and surface arcs. (a)-(b) The dispersions of the surface state with fixed $k_z = 0.5\,\pi/h$ on the N and H surfaces, respectively. (c)-(f) The equi-frequency contours of the surface state at fixed frequency $f = 7\text{kHz}$ and $f = 9\text{kHz}$ on the N and H surfaces, respectively. The color maps denote the experimental data, and the white dashed and black lines represent the calculated dispersions of surface state and projected bulk state, respectively.

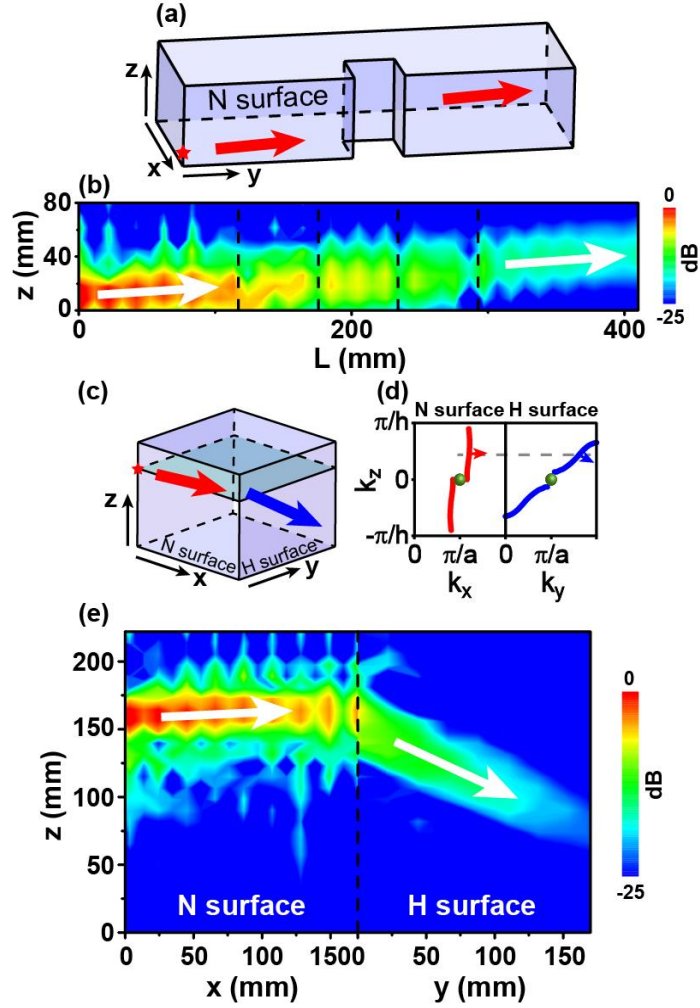

**Figure 4.** Experimental topological reflection-free propagations of acoustic surface arc wave. (a) Schematic of a sample with the multiple joints of adjacent facets on the N surface. (b) Experimental robust propagation of surface arc wave through the multiple joints of adjacent facets. (c) Schematic of topological negative refraction at the interface between the N and H surfaces. (d) Group velocities (red and blue arrows) of the surface arc wave of the N and H surfaces. The non-closed properties of the surface arcs guarantee the reflection immunity. (e) Experimental observation of topological negative refraction. In both the two experiments of wave propagation, the field distributions are measured at the frequency 7kHz. The red stars denote the position of excitation, and the arrows represent the propagating directions.